\documentclass[twocolumn,showpacs,preprintnumbers,amsmath,amssymb]{revtex4}

\usepackage{graphicx}
\usepackage{dcolumn}
\usepackage{bm}
\usepackage{mathrsfs}
\usepackage{color}

\begin{document}

\title{
Coulomb Energy of $\alpha$-Aggregates on a Soap Bubble Shape
}%

%
%
%
\author{Akihiro Tohsaki}

\affiliation{
Research Center for Nuclear Physics (RCNP), Osaka University,
10-1 Mihogaoka, Ibaraki, Osaka 567-0047, Japan
}

\author{Naoyuki Itagaki}

\affiliation{
Yukawa Institute for Theoretical Physics, Kyoto University, 606-8502 Kyoto, Japan 
}

\date{\today}

\begin{abstract}
We study the property of $\alpha$-aggregates
on a soap bubble shape within a microscopic framework, which takes full account of the Pauli principle.
Our special attention is payed to the Coulomb energy for such an exotic shapes of nuclei,
and we discuss  the advantage 
of $\alpha$-clusters with geometric configurations
compared with the uniform density distributions in reducing the repulsive effect.
We consider four kinds of configurations of $\alpha$
clusters on a soap bubble, which are dual polyhedra composed of a dodecahedron and an icosahedron,
octacontahedron and two types of truncated icosahedrons, that is, two kinds of Archimedean solids.
The latter two are an icosidodecahedron and a fullerene shape. When putting each $\alpha$-cluster
on the vertex of polyhedra, four $\alpha$-cluster aggregates correspond to the following four nuclei;
Gd (64 protons), Po (84 protons), Nd (60 protons) and a nucleus with 120 protons, respectively.
\end{abstract}

\maketitle

\section{Introduction}

This report is written as a series of the study 
on $\alpha$-cluster structure of heavy nuclei by microscopic aspect.
We have already examined the extreme case of hollow configurations and
pointed out that the systems have clear energy minimum points when $\alpha$ clusters approach from large distances,
and we found that such shapes help very much in reducing the Coulomb repulsion~\cite{PhysRevC97.011301R}.
For such studies,
it is also inevitable to quantitatively scrutinize the effect of the Pauli principle with respect to the Coulomb energy, because 
the contribution of the exchange term works attractively. 

The Coulomb energy is one of the main players against the stability of heavy nuclei owing to its strong repulsion. 
We know that the Coulomb energy is essential in nuclear fission, and also in the nuclear structure,
whose quantity appreciably depends on the configuration of the protons. 
For instance, according to the electrostatics, the Coulomb energy of uniform density of positive point-charges inside a sphere is more repulsive than that on its surface as if they float on a soap bubble. Namely, the former is ${3 \over 5}{Q^2 \over \rho}$, the latter is  ${1 \over 2}{Q^2 \over \rho}$ , where $Q$ is total charge and $\rho$ is the radius of the sphere.  
This suggests the possibility that contribution of nonuniform distribution contributes in reducing the Coulomb repulsion.
A long standing history of the studies for the
shape of heavy nuclei has suggested the existence of special shape called as thin spherical shell
nuclei~\cite{Wilson}, bubble nuclei~\cite{Bulgac}, or torus nuclei~\cite{Wong,Ichikawa}, which prevent drastically the Coulomb repulsion.
An appreciable dent in the middle of density distribution of $^{208}$Pb  has been observed, which may support 
the mixing of components of hollowing nuclei~\cite{Heisenberg}.

For the proton distribution, it
is natural to consider that 
they are in $\alpha$ clusters, since $\alpha$ clusters are the most stable existence 
compared with the other nuclear clusters.  
With this assumption,
it is important to seek for the optimum configurations of
$\alpha$-clusters for heavier nuclear systems.

We again take a microscopic
$\alpha$-cluster model in Brink-Bloch parameter space~\cite{Brink} 
and focus on the relation between the 
geometric configurations of $\alpha$ clusters and Coulomb repulsion more quantitatively.
Here geometrical configurations are assumed
within the framework, which takes full account of the Pauli principle. 
We can discuss the Coulomb energy with the Pauli principle extracting the corresponding part from the total binding energy.
Therefore, in this report, we focus only upon the Coulomb energy for $\alpha$-clusters on a soap bubble shape
and show how they are favored from Coulomb energy point of view.
 
As examples, we take four kinds of configurations originating in the icosahedron.
\begin{itemize}
\item[1.]
An Archimedean solid (truncated polyhedron); the 30 $\alpha$ clusters are put on each center of 30 edges of the icosahedron (corresponding to Nd).
\item[2.]
A dual polyhedron composed of dodecahedron and icosahedron;  the 20
$\alpha$-clusters are put on each center of 20 surfaces and 12 $\alpha$-clusters on the 12 vertexes of the icosahedron (corresponding to Gd).
\item[3.]
An octacontahedron; the 12 $\alpha$ clusters are put on the 12 vertexes and 30 $\alpha$ clusters on each center of the 30 edges of the icosahedron  (corresponding to Po).
\item[4.]
Another Archimedean solid, the 60 $\alpha$ clusters have a well known fullerene shape.
\end{itemize}
Note that the second and third cases are adjusted to put all the $\alpha$ clusters on the same sphere (with the radius parameter $\rho$)
and they correspond to nuclei, Nd, Gd, Po and the unknown super heavy nucleus with $Z=120$.  
In Fig. 1, we show the schematic features of four cases, 
(a): 30 $\alpha$'s (Nd), (b): 32 $\alpha$'s 
(Gd), (c): 42 $\alpha$'s, and (d): 60 $\alpha$'s (fullerene). 
Here the red balls mean the $\alpha$-clusters on the vertexes of the icosahedron (Fig. 1 (b), (c)), and blue balls come from $\alpha$ clusters on the edges of icosahedron
(Fig. 1 (a), (c), (d)).
In Fig. 1 (b), the blue balls correspond to the $\alpha$ clusters on the surfaces of the icosahedron.
All the $\alpha$ clusters are on the same sphere with the radius parameter $\rho$.

\begin{figure}[t]
	\centering
	\includegraphics[width=9.cm]{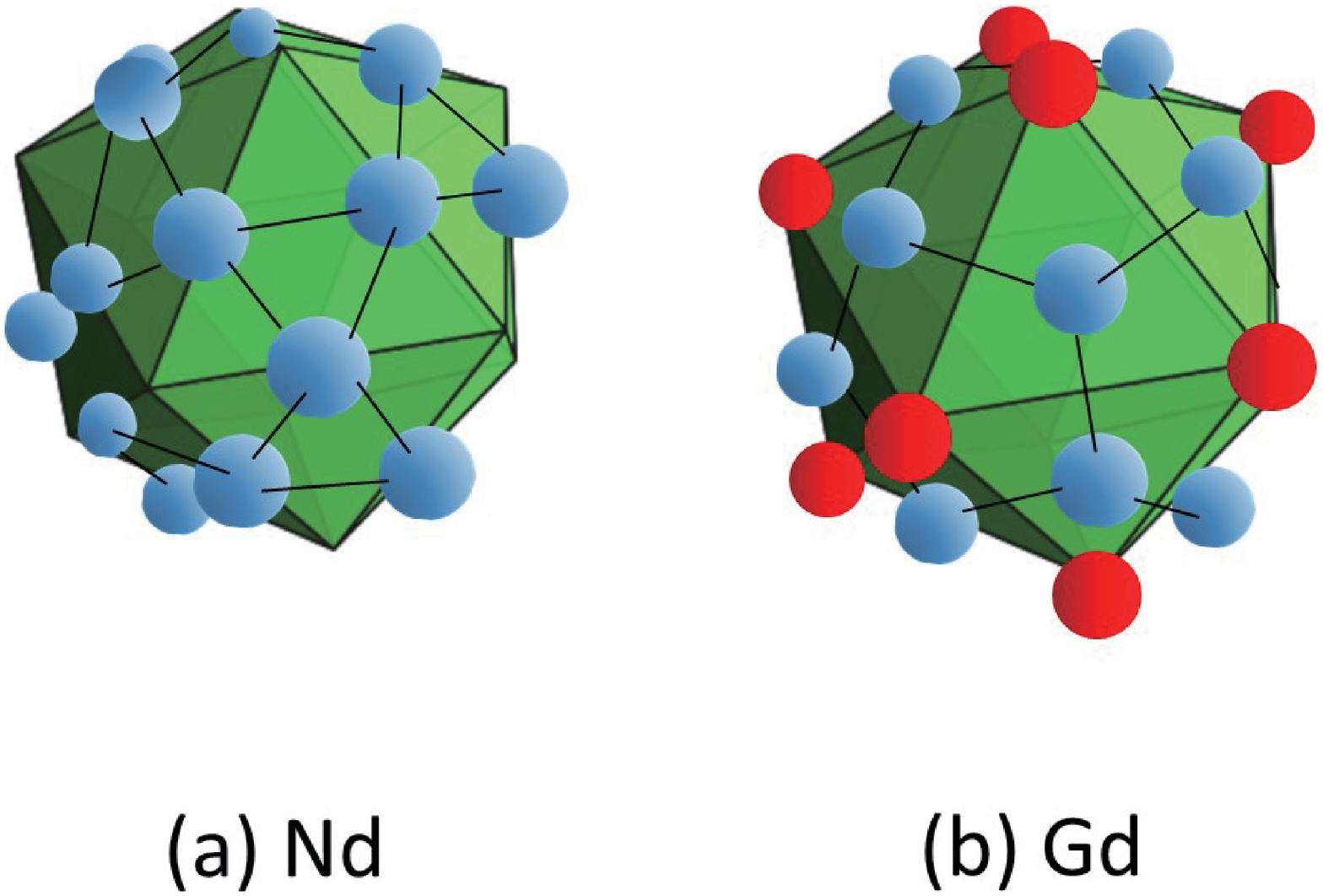} 
	\includegraphics[width=9.cm]{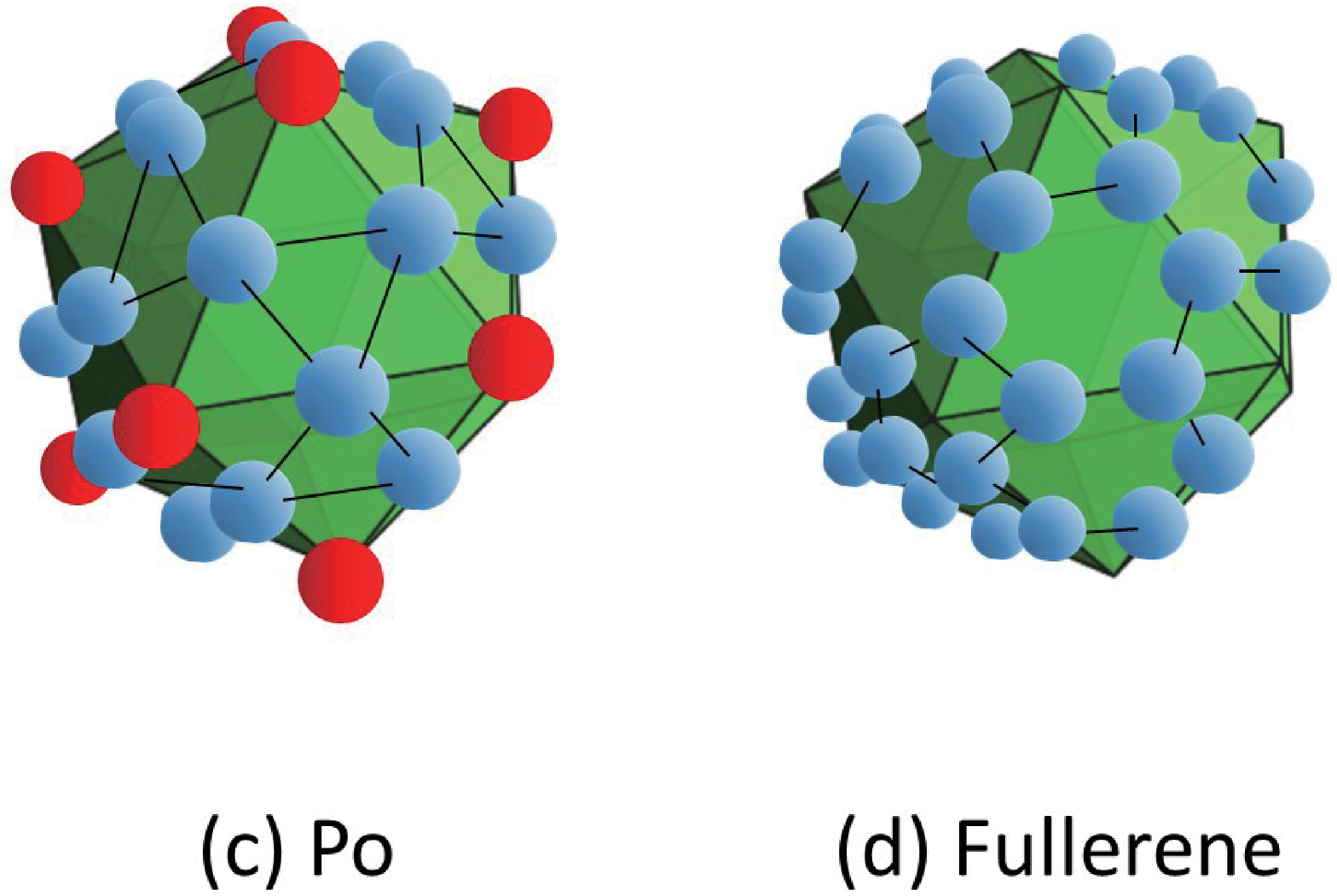} 
	\caption{
Schematic figure for introduced configurations, .(a): 30 $\alpha$'s (Nd), (b): 32 $\alpha$'s 
(Gd), (c): 42 $\alpha$'s, and (d): 60 $\alpha$'s (fullerene).
The red and blue balls show $\alpha$ clusters. 
The red balls mean the $\alpha$-clusters on the vertexes of the icosahedron ((b), (c)), 
and blue balls come from $\alpha$ clusters on the edges of icosahedron
((a), (c), (d)).
In (b), the blue balls correspond to the $\alpha$ clusters on the surface of icosahedron.
     }
\label{schematic}
\end{figure}

Fuller intuitively showed that various kinds of structural architectures stably have a hollow inside. Their origin is in an icosahedral skeleton~\cite{Edmondson}. 
In an atomic world, such a structure already appears 
as the fullerene~\cite{Kroto,Kratschmer} and 
nano-tubes~\cite{Iijima-1,Iijima-2}. 
He insists on the interplay between physical forces and spatial constraints, which guarantee geometrical structure of architectures based on tetrahedra and icosahedra.  In nuclear physics, in addition to them, the Pauli principle also plays an essential role in the requirement of minimum energy of nuclei. 
The dual role of the Pauli principle, which acts repulsively when two $\alpha$ clusters approach with each other,
and which gives an attractively effect for separated two $\alpha$ clusters, is regarded as another kind of the
spatial constraint based on the quantum mechanics~\cite{Tamagaki-1,Tamagaki-2}.
We hope to see analogous structures in a nucleonic world after clarifying the role of neutrons, which may provide a stability against the Coulomb repulsion.
In other word, we propose a quantum mechanical $\alpha$ cluster architecture 
based on full microscopic quantum mechanics. 

\section{Formulation \label{model}}

We adopt an $\alpha$-clustering standpoint with a microscopic framework. We extract only the Coulomb energy from the total binding energy including kinetic, effective inter-nucleon force. We, here, employ Brink-Bloch type wave function for $n\alpha$ clusters, which takes full account of the Pauli Principle:

\begin{equation}
\Psi (\rho) = {\cal A}
\{
\phi(\rho {\bm R_1})
\phi(\rho {\bm R_2})
\cdots
\phi(\rho {\bm R_n})
\},
\label{total-wf}
\end{equation}
where ${\cal A}$ is the antisymmetrization operator among all the nucleons.
The $n\alpha$ clusters are on the surface of the sphere with the radius $\rho$ (fm),
and
the vectors ${\bm R}_1, {\bm R}_2, \ldots {\bm R}_n$  are the parameters on the dimensionless unit sphere as shown in Fig. 1.
The $k$-th $\alpha$ cluster ($k = 1, 2, \cdot \cdot n$) wave function is written by
\begin{equation}
\phi(\rho {\bm R_k}) =  \prod_{i,j=1,2}
\left(\frac{1}{\pi b^2} \right)^{\frac{3}{4}}
\exp \{- {1 \over 2b^2} \left({\bm r}^{ij}_k - \rho {\bm R}_k \right)^{2} \} \chi^{ij}_k,
\label{Brink-wf}
\end{equation}
where $b$ is the nucleon size parameter, and $\chi^{ij}_k$ is a spin isospin wave function.
The vector ${\bm r}^{ij}_k$ is the real physical coordinate for the nucleon,
and $i$ and $j$ are labels for the spin and isospin, respectively,
for the four nucleons in the $k$-th $\alpha$ clusters.
The four nucleons in the $k$-th $\alpha$ cluster share the common Gaussian center,
$\rho {\bm R}_k$.
We prepare four sets of $\{ \bm R_1, \cdots, \bm R_N \}$
corresponding to the configurations in Fig.~1.
The norm and energy kernel matrix elements
after carrying out the integration with respect to the real physical coordinates
$\{ {\bm r}^{ij}_k \}$ are functions of variational parameter $\rho$.
The Coulomb energy operator is written by
\begin{eqnarray}
V^{(c)} =
{1 \over 2} \sum_{k,l,i,i'}
{e^2 \over |{\bm r}^{i1}_k - {\bm r}^{i'1}_l|},
\label{Coul}
\end{eqnarray}
which acts only on the terms with $j$=1, namely, on the protons.
Thus the Coulomb energy ($E_c(\rho)$) is defined by
\begin{equation}
E_c(\rho) =
{
\langle \Psi (\rho) |V^{(c)}| \Psi (\rho) \rangle
\over
\langle \Psi (\rho) | \Psi (\rho) \rangle.
}
\label{full-c}
\end{equation}
The numerator is given by
\begin{eqnarray}
&& \langle \Psi (\rho) |V^{(c)}| \Psi (\rho) \rangle \nonumber \\
&&
= \langle \Psi (\rho) | \Psi (\rho) \rangle \nonumber \\
&& \times
\sum_{ii'kk'll'}
\langle \phi(\rho {\bm R_k}) \phi(\rho {\bm R_l})
| {e^2 \over |{\bm r}^{i1}_{k'} - {\bm r}^{i'1}_{l'}|}  |
\phi(\rho {\bm R_{k'}}) \phi(\rho {\bm R_{l'}})  \rangle \nonumber \\
&& \times
(2G^{-1}_{k'k}G^{-1}_{l'l}-G^{-1}_{k'l}G^{-1}_{l'k}),
\label{d-e}
\end{eqnarray}
which is the sum of direct (proportional to $G^{-1}_{k'k}G^{-1}_{l'l}$)
and exchange (proportional to $G^{-1}_{k'l}G^{-1}_{l'k}$) terms,
where $G_{kk'}$ makes $n \times n$ matrix of which each element is given by
\begin{eqnarray}
G_{kk'} \equiv &&  \langle \phi(\rho {\bm R_k}) | \phi(\rho {\bm R_{k'}}) \rangle \nonumber \\
       = && \exp \{-{ \rho^2 \over 4b^2} ( {\bm R_k} - {\bm R_{k'}})^2 \}.
\end{eqnarray}

In Eq.~(4), the internal Coulomb energy of proton-pairs in $\alpha$-clusters is inevitably included as $nE_{in}$.
The each term for the Coulomb energy operator is given by an analytical form:
\begin{eqnarray}
&& \langle \phi(\rho {\bm R_k}) \phi(\rho {\bm R_l})
| {e^2 \over |{\bm r}^{i1}_{k'} - {\bm r}^{i'1}_{l'}|}  |
\phi(\rho {\bm R_{k'}}) \phi(\rho {\bm R_{l'}})  \rangle   \nonumber \\
&& = G_{kk'} G_{ll'} {2e^2 \over s_{klk'l'}} {\rm erf}({1 \over 2} s_{klk'l'}),
\end{eqnarray}
where
\begin{equation}
s_{klk'l'} = {1 \over \sqrt{2} b} \rho | {\bm R}_k - {\bm R}_l + {\bm R}_{k'} - {\bm R}_{l'} |,
\end{equation}
where erf is an error function.
We should compare the results with those from which the antisymmetrization operator is switched off. Namely, the term without the Pauli principle, so-called direct term, is symbolically denoted by $E_{c(d)} (\rho)$.  Note that the Coulomb energy in microscopic model
depends on the
size parameter of $b$ in Eq.~\eqref{Brink-wf}.
We are also interested in the comparison of results with
those of point-charged approximation of each $\alpha$. The electrostatics
teaches us the results:
\begin{eqnarray}
E_{pc} (\rho) = &&
 {4e^2 \over \rho} 
\sum_{k < l}^n 
{1 \over |{\bm R}_k - {\bm R}_l| }
+ nE_{in} \nonumber \\
= && C_s(n) {4e^2 \over \rho} n(n-1) + nE_{in},
\label{pointc}
\end{eqnarray}
where $n$ is number of $\alpha$ clusters and $C_s(n)$ is a constant with respect to the geometric
configuration of $\alpha$ clusters as shown in Table I.
We should point out that these values approach 1/2
corresponding to that of uniform distribution of positive
particles on a soap bubble in limiting case of $n \to \infty$. We,
here, know that even such values directly depend on the
$\alpha$-cluster configurations. In addition to three quantities
on the Coulomb energy, $E_c(\rho)$, $E_{c(d)}(\rho)$, and $E_{pc}(\rho)$, 
we
compare two cases, 
\begin{equation}
E_{c(uS)}(\rho) = {1 \over 2} {4e^2 \over \rho} n^2  +nE_{in},
\label{unisu}
\end{equation}
and 
\begin{equation}
E_{c(uV)} (\rho) =
{3 \over 5}{4e^2 \over \rho} n^2+nE_{in},
\label{unisp}
\end{equation} 
which are those of uniform $\alpha$ cluster distribution on the
surface of the soap bubble and uniform $\alpha$ cluster distribution inside the sphere. 
The total charge
$Q$ is given by $2ne$. In the following section we discuss
these five numerical values for four configurations,
where the contribution of the Coulomb energy is divided 
by $n$ (half of proton numbers).

\begin{table}
\caption{The $C_s(n)$ values for the point-charged configurations
defined in Eq.~\eqref{pointc}, where $n$ stands for the number of $\alpha$ clusters.
}
  \begin{tabular}{c|c} \hline \hline
 $n$ & $C_s(n)$    \\ \hline
   30  &   0.4153\\
   32  &   0.4156 \\
   42  &   0.4252 \\ 
   60  &   0.4384   \\\hline
  \end{tabular}
\end{table}

\section{NUMERICAL RESULTS AND DISCUSSION}

In Figs. 2$\sim$5, we show the Coulomb energy for one $\alpha$ cluster versus the
radius $\rho$ for the four cases, Nd (Fig.~2), Gd (Fig.~3), Po (Fig.~4), and Fullerene (Fig.~5). 
Here, thick-solid, solid, dotted, dashed, and dash-dotted lines 
represent the Coulomb energy for one $\alpha$
 defined in Eq.~\eqref{full-c} ($E_c(\rho)/n$), 
its direct term ($E_{c(d)}(\rho)/n$), 
point charge approximation defined in Eq.~\eqref{pointc} ($E_{pc} (\rho)/n$),
uniform $\alpha$ cluster distribution on a surface defined in Eq.~\eqref{unisu}
($E_{c(uS)}(\rho)/n$),
and 
uniform $\alpha$ cluster distribution in a sphere defined in Eq.~\eqref{unisp}
($E_{c(uV)}(\rho)/n$),
respectively.
For the number of neutrons, the Coulomb energy is common
for all the isotopes of an individual atom. However, the
appropriate $\rho$ depends on the radius of isotopes with the
same proton number, which are related to the neutron
number $N$. In this model, the main part of the neutrons
is trapped in the $\alpha$-clusters, then the excess of neutrons is
estimated as $N-2n$. If the excess neutrons exist inside,
all the protons in the $\alpha$-clusters 
float on the sphere. On
the other hand, they are outside, then the nucleus has an
appreciable cavity. Anyhow, the radius $\rho$ of the sphere
is not independent of the number of excess neutrons, in
assuming that the nuclear radius is the function of only
the mass number $A = Z + N$. The possible range of $\rho$ is
shown in Figs. 2$\sim$5 by a belt with net in assuming the
nuclear radius $r_0 A^{1/3}$ with $r_0 = 1.2$ fm and 
plausible mass numbers,
$2.0 Z \leq A \leq 2.7 Z$, 
where for heavy nuclei, not only many isotopes
but also a variety of positions of $\alpha$-cluster are imagined.

\begin{figure}[t]
	\centering
	\includegraphics[width=6.5cm]{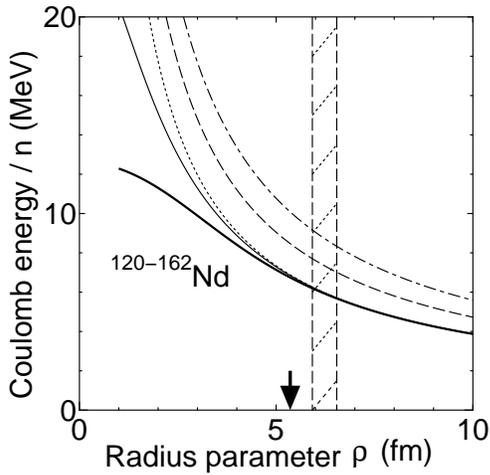} 
	\caption{
Coulomb energy for one $\alpha$ for Nd isotopes.
Thick-solid, solid, dotted, dashed, and dash-dotted lines 
represent the results of 
$E_c(\rho)/n$, $E_{c(d)}(\rho)/n$, $E_{pc} (\rho)/n$, $E_{c(uS)}(\rho)/n$, and $E_{c(uV)}(\rho)/n$,
respectively.
The arrow at $\rho = 5.3$ fm gives 
the nearest neighbor $\alpha$-$\alpha$ distance of 3.3 fm,
which is the optimal value in the free space.
     }
\label{nd-coul}
\end{figure}

\begin{figure}[t]
	\centering
	\includegraphics[width=6.5cm]{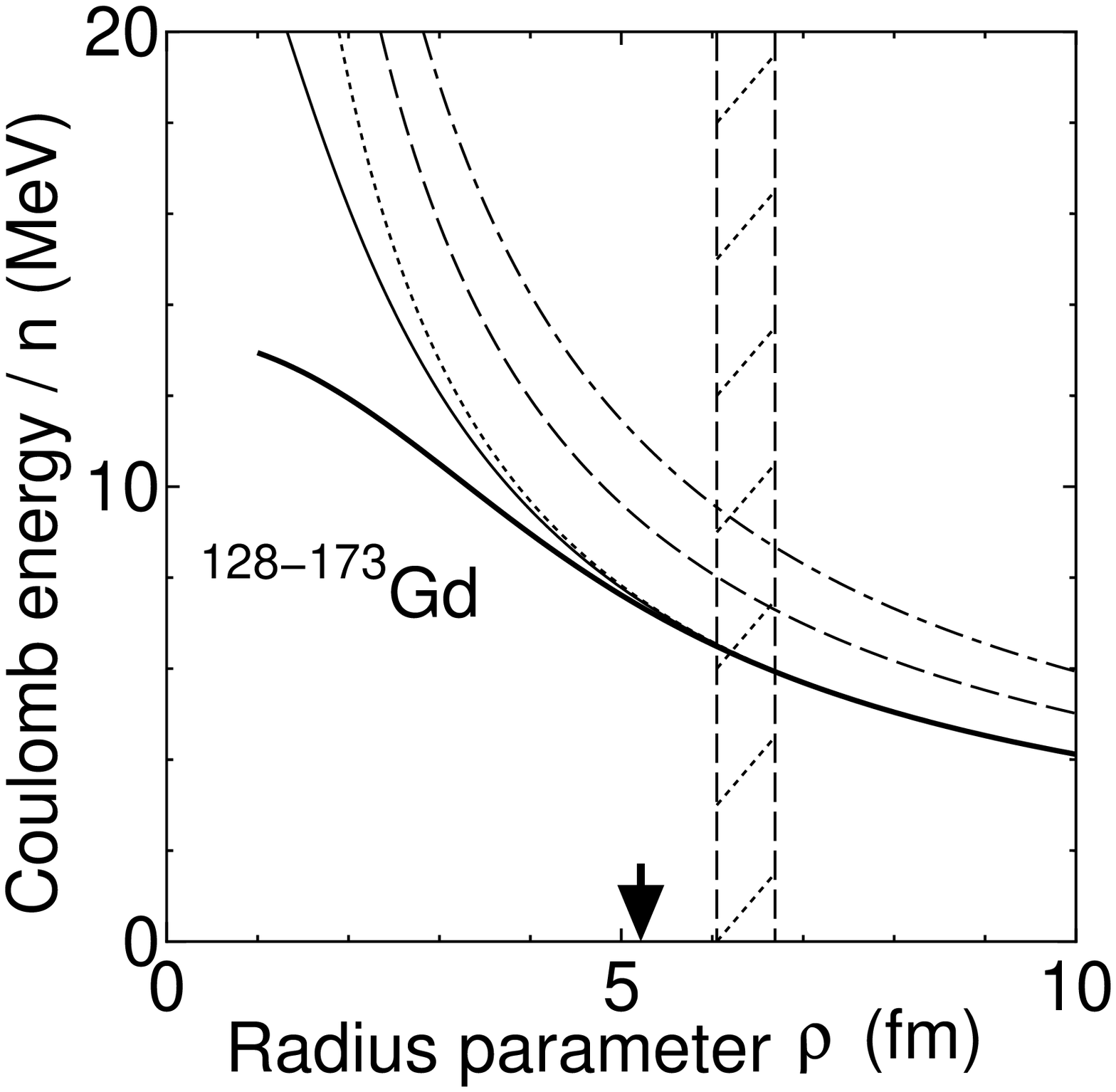} 
	\caption{
Coulomb energy for one $\alpha$ for Gd isotopes.
Thick-solid, solid, dotted, dashed, and dash-dotted lines 
represent the results of 
$E_c(\rho)/n$, $E_{c(d)}(\rho)/n$, $E_{pc} (\rho)/n$, $E_{c(uS)}(\rho)/n$, and $E_{c(uV)}(\rho)/n$,
respectively.
The arrow at $\rho = 5.2$ fm gives 
the nearest neighbor $\alpha$-$\alpha$ distance of 3.3 fm,
which is the optimal value in the free space.
     }
\label{gd-coul}
\end{figure}

\begin{figure}[t]
	\centering
	\includegraphics[width=6.5cm]{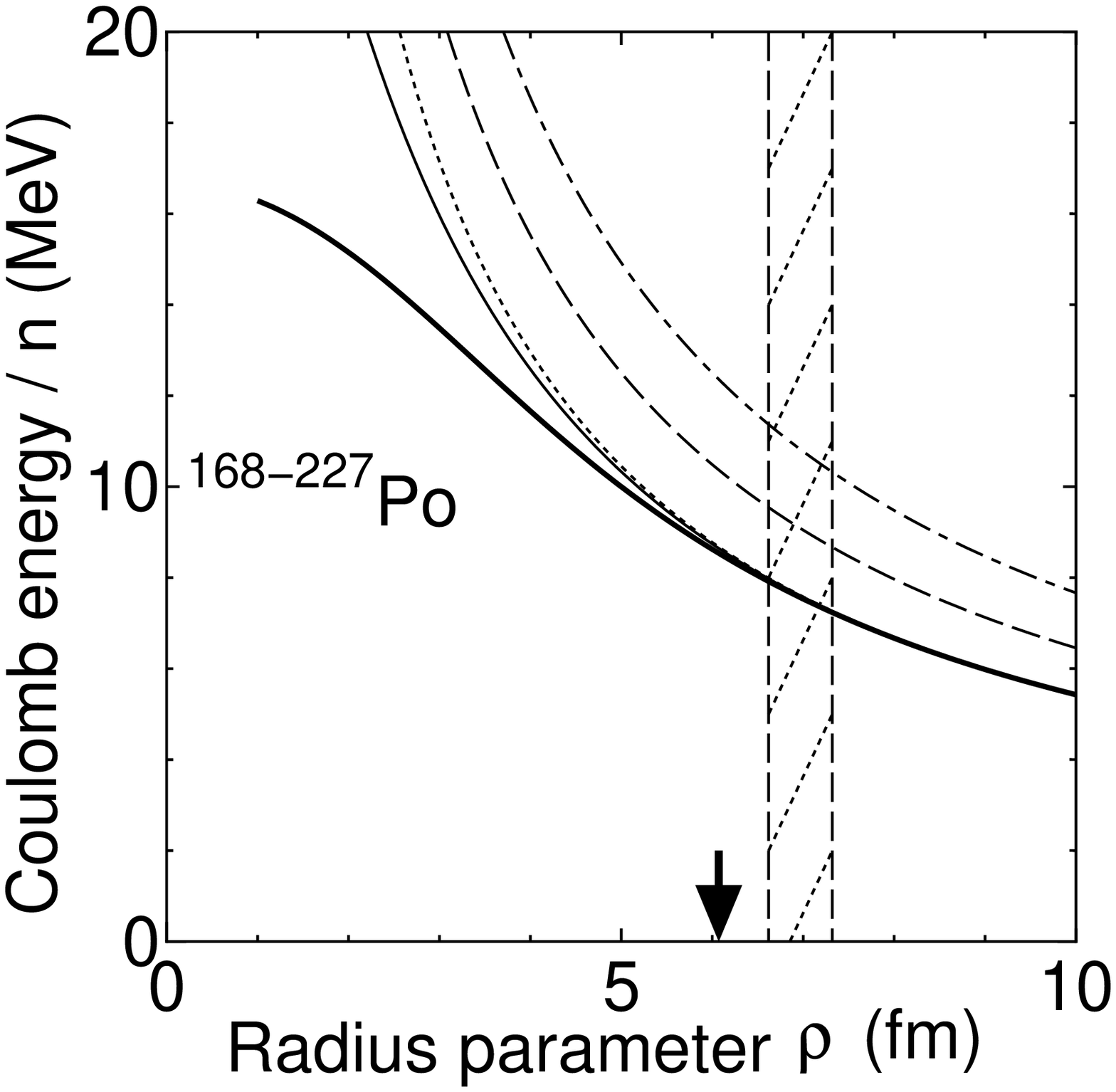} 
	\caption{
Coulomb energy for one $\alpha$ for Po isotopes.
Thick-solid, solid, dotted, dashed, and dash-dotted lines 
represent the results of 
$E_c(\rho)/n$, $E_{c(d)}(\rho)/n$, $E_{pc} (\rho)/n$, $E_{c(uS)}(\rho)/n$, and $E_{c(uV)}(\rho)/n$,
respectively.
The arrow at $\rho = 6.1$ fm gives
the nearest neighbor $\alpha$-$\alpha$ distance of 3.3 fm,
which is the optimal value in the free space.
     }
\label{po-coul}
\end{figure}

\begin{figure}[t]
	\centering
	\includegraphics[width=6.5cm]{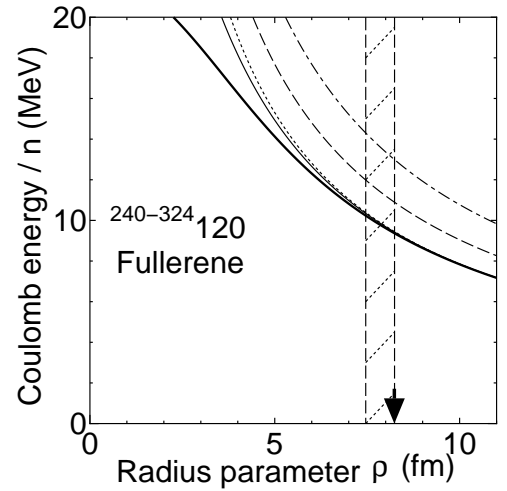} 
	\caption{
Coulomb energy for one $\alpha$ for Po isotopes.
Thick-solid, solid, dotted, dashed, and dash-dotted lines 
represent the results of 
$E_c(\rho)/n$, $E_{c(d)}(\rho)/n$, $E_{pc} (\rho)/n$, $E_{c(uS)}(\rho)/n$, and $E_{c(uV)}(\rho)/n$,
respectively.
The arrow at $\rho = 8.2$ fm gives 
the nearest neighbor $\alpha$-$\alpha$ distance of 3.3 fm,
which is the optimal value in the free space.
     }
\label{po-coul}
\end{figure}

It should be noted that even for heavy nuclei, in
cluster models, we can correctly obtain numerical values
without any round-off errors in the 
radius parameter region of $1.5 \ {\rm fm} \geq \rho$. 
Although the norm kernel, $\langle \Psi (\rho) | \Psi (\rho) \rangle$, 
is incredibly
small in the small radius of the sphere, the inverse matrix
of $G_{ik}$ is not divergent.
We can
indicate many interesting features from Figs. 2$\sim$5 as
follows:

\begin{itemize}
\item[1.] The Pauli principle drastically influences the region with small
$\rho$ values of the sphere and decreases the energies in all nuclei. 
We can know, unlike the curves without the Pauli principle (solid lines), 
a definite convergence of the Coulomb energy in the case of thick solid lines at small 
$\rho$ regions.
The exact Coulomb energy in thick solid lines converged to finite values at  $\rho \to 0$, 
which cannot be seen in other lines.
This effect works in reducing the incompressibility of heavy nuclei.
\item[2.] On the contrary, in the range of the belt with net corresponding 
to plausible radii of the isotopes 
(the left and right edges correspond to 
$\rho_l = r_0(2.0Z)^{1/3}$
and $\rho_r$ = $r_0(2.7Z)^{1/3}$, respectively),
three curves,
$E_c(\rho)/n$ (thick solid lines), $E_{c(d)}(\rho)/n$ (solid lines), and $E_{pc}(\rho)/n$ (dotted lines), 
almost coincide with each other in all figures. 
Surprisingly, even the point-charge
model well works in this region. 
This means that
the point charged cluster model, where the Pauli principle is switched off,
may be useful for the studies of heavy nuclei based on the cluster models
as the first step.
For instance, it is possible to employ Ali-Bodmer
force between two $\alpha$-clusters~\cite{AliBodmer}
and phenomenological $N$-$\alpha$ force~\cite{Satchler}.
\item[3.] The curves with geometric $\alpha$ cluster configurations, $E_c(\rho)/n$ (thick solid lines),
$E_{c(d)}(\rho)/n$ (solid lines), and $E_{pc}(\rho)/n$ (dotted lines), are quite different from that
of the uniform density distribution of $\alpha$ clusters 
on the surface ($E_{c(uS)}(\rho)/n$, dashed lines) 
and
that in the sphere ($E_{c(uV)}(\rho)/n$, dash-dotted lines)
in all figures. 
Therefore, the $\alpha$ clusters with geometric configurations
may be responsible for the study of heavy nuclei, when they are on the surface.
\item[4.] The mixing of geometrical configurations drastically reduces 
the Coulomb repulsion, and
as the dent
in the middle of $^{208}$Pb  has been observed~\cite{Heisenberg},
the uniform distribution of protons, the case of
$E_{c(uV)}(\rho)/n$ may not be plausible, which is a basic idea
of Bethe-Weizsaecker's mass formula. 
This fact
may deeply require understanding the distribution
of nucleons in nuclei from scratch.
\end{itemize}

\begin{table}
\caption{The $p(\rho)$ values for the Pauli effect defined in Eq.~\eqref{Pauli},
and $\rho_l = r_0(2.0Z)^{1/3}$ 
and $\rho_r$ = $r_0(2.7Z)^{1/3}$ 
correspond to the left and right edges, respectively.
The quantities $d_l$ and $d_r$ 
mean the nearest neighbor distance of $\alpha$-$\alpha$ 
on the surface
obtained with $\rho = \rho_l$ and $\rho_r$, respectively. 
}
  \begin{tabular}{c|ccc|ccc} \hline \hline
 $n$   & $\rho_l$ (fm) & $p(\rho_l)$ & $d_l$ &  $\rho_r$ & $p(\rho_r)$&  $d_r$ \\ \hline
   30  &   5.9    & 0.984  & 3.7         & 6.5 & 0.870 & 4.1  \\
   32  &   6.1    & 0.947  & 3.9         & 7.0 & 0.578 & 4.5  \\
   42  &   6.6    & 0.998  & 3.6         & 7.3 & 0.939 & 4.0  \\ 
   60  &   7.5    & 1.000  & 3.0         & 8.2 & 1.000 & 3.3  \\\hline
  \end{tabular}
\end{table}

We consider the effect of the Pauli principle with respect to the radius 
$\rho$ of the sphere. The diagonal part of
the norm kernel has the following property depending on
the Pauli principle:

\begin{equation}
\lim_{\rho \to 0} \langle \Psi (\rho) | \Psi (\rho) \rangle = 0,
\end{equation}
and
\begin{equation}
\lim_{\rho \to \infty} \langle \Psi (\rho) | \Psi (\rho) \rangle = 1.
\end{equation}
When we consider only the proton contribution, the
index of the Pauli effect is taken as
\begin{equation}
p(\rho) = 1 - \sqrt{ \langle \Psi (\rho) | \Psi (\rho) \rangle }.
\label{Pauli}
\end{equation}
The effect coming from the off-diagonal part (different $\rho$ for bra and ket states) 
is not considered, which should be taken into account when
studying these nuclei dynamically.
At both sides
of the edges of the belt with net in Fig.~2$\sim$5
(the left and right edges correspond to 
$\rho_l = r_0(2.0Z)^{1/3}$ 
and $\rho_r$ = $r_0(2.7Z)^{1/3}$, respectively), 
we give the quantities
of $p(\rho)$ in Table II. 
The quantities $d_l$ and $d_r$ 
mean the nearest neighbor distance
between two $\alpha$-clusters on the surface of the spheres
obtained with $\rho = \rho_l$ and $\rho_r$, respectively. 
We see from Table II appreciable effect in this region. 
Nevertheless the three kinds of the Coulomb energies 
($E_c(\rho)/n$, $E_{c(d)}(\rho)/n$, $E_{pc} (\rho)/n$)
well
coincide with each other as mentioned before, thus we need detailed analysis 
on the role of
the Pauli principle, which can be described in terms of the
exchange number of nucleons. 


Anyhow, it is natural that the heavier the nucleus is, the
stronger the Coulomb energy is. Thus, we should
study the stability of $Z = 120$, for instance, by applying an effective inter-nucleon force appropriate
for the cluster model to anneal the Coulomb repulsion.
We are also waiting for not only the next research taking account
of the effective inter-nucleon force but also the consideration of the
oozy of excess neutrons.

Our previous report 
obtained by applying Tohsaki F1 force~\cite{F1-force} as inter-nucleon force
has pointed out that the property of 
$\alpha$-$\alpha$ interaction remains even inside heavy nuclei,
such as 60 $\alpha$ clusters with a fullerene shape~\cite{PhysRevC97.011301R}.
Namely, the relative distance of nearest neighbor of $\alpha$-$\alpha$ is  always around 3.3 fm,
which is the optimal distance between $\alpha$ clusters in the free space.
In Figs.2$\sim$5, thick arrows show the radius of sphere $\rho$, 
in which nearest neighbor corresponds to 3.3 fm. 
Here the place of arrows for comparably light nuclei (Nd, Gd, Po) is smaller than the belt, 
on the other hand, the case with $Z=120$, the arrow is inside the belt. 
As for former three cases, the excess neutrons, which are not contained in $\alpha$ clusters, 
may widely exist outside of the sphere. On the other hand, 
unknown ultra super heavy nucleus with $Z=120$ has 60 $\alpha$-clusters floating 
on surface of the sphere, which enfolds excess neutrons. 
In Table III, five kinds of energy quantities 
($E_c/n$, $E_{c(d)}/n$, $E_{pc}/n$, $E_{c(uS)}/n$, and  $E_{c(uV)}/n$)
are listed for the fixed nearest neighbor distance with 
$d = 3.3$ fm.
We may point out that the former there nuclei, which have arrow positions before the belt regions, have appreciable quantity of the Pauli principle, 
where the $E_c/n$ values are slightly lower than $E_{c(d)}/n$.

\begin{table}
\caption{The energies of   
$E_c/n$, $E_{c(d)}/n$, $E_{pc}/n$, $E_{c(uS)}/n$, and  $E_{c(uV)}/n$ (all in MeV) 
for the four geometric configurations shown in Fig.~1, where $n$ is number of $\alpha$ cluster.
The energies are calculated at the radius parameter $\rho$, which gives 
3.3 fm for the nearest neighbor distance. 
The $\alpha$-$\alpha$ distance of 3.3 fm is the optimal one in the free space.
}
  \begin{tabular}{c|cccccc} \hline \hline
 $n$   & $\rho$  & $E_c/n$ & $E_{c(d)}/n$ & $E_{pc}/n$ & $E_{c(uS)}/n$ &  $E_{c(uV)}/n$ \\
       &  (fm)   &    (MeV)      &              (MeV) &             (MeV) &               (MeV) & (MeV) \\ \hline
   30  &   5.34   & 6.76  & 6.87 & 6.90 &  8.50 & 10.11 \\
   32  &   5.15   & 7.44  & 7.57 & 7.61 &  9.35 & 11.14 \\
   42  &   6.04   & 8.56  & 8.69 & 8.72 & 10.42 & 12.42 \\ 
   60  &   8.18   & 9.43  & 9.49 & 9.51 & 10.87 & 13.08 \\\hline
  \end{tabular}
\end{table}

\section{SOME REMARKS}
Following our previous report~\cite{PhysRevC97.011301R}, we studied the Coulomb
energy of the $\alpha$-clusters on a soap bubble. We, here,
indicated the advantage of the geometric configurations
of $\alpha$-clusters in reducing the Coulomb repulsion by comparing with two types of uniform 
distributions of $\alpha$ clusters, namely, an uniform distribution 
inside the sphere and that
on the surface of the sphere. 
We also have shown that the Pauli principle drastically influences the region with small
radius parameter $\rho$ of the sphere and decreases the energies in all nuclei. 
A definite convergence of the Coulomb energy has bee shown
at  $\rho \to 0$
unlike the curves without the Pauli principle. 
On the contrary, in the range of plausible radii of the isotopes 
even the point-charge
model well works. 
This means that
the point charged cluster model, where the Pauli principle is switched off,
may be useful for the studies of heavy nuclei based on the cluster models
as the first step.

However, in this model, it
is indispensable to investigate the distribution of the 
excess neutrons. Uniform distribution of excess neutrons,
otherwise di-neutron pairing, and inside or outside, there
are various kinds of possibilities. Before that, we should
find out the most reliable inter-nucleon force including
many-body terms for cluster model due to the guarantee
of the saturation property of nuclear matter. 
In our previous study for the geometric configurations~\cite{PhysRevC97.011301R}, 
we have utilized Tohsaki F1 force~\cite{F1-force}, which guarantees
the saturation properties and also reproduces the $\alpha$-$\alpha$ 
scattering phase shift. 
This interaction should be more examined in neutron-rich side,
as we have introduced for light neutron-rich nuclei~\cite{C12,O24}.
In order
to step in the world of fundamental phenomena in heavy
nuclei, which contain $\alpha$-decay, $\beta$-decay, fission and so on,
it is necessary for us to clarify the role of excess neutrons within
the microscopic cluster model. This is because the microscopic cluster model, 
which naturally includes the ground
state of the shell model, can exactly evaluate the Pauli
principle.

In this article, we discussed the geometric configurations;
however the opposite aspect is gas-like behavior of the $\alpha$ clusters.
We have previously introduced 
the Tohsaki Horiuchi Schuck R\"{o}pke (THSR) wave function 
for the studies of gas-like nature of $\alpha$ clusters
in various nuclei including the so-called Hoyle state of $^{12}$C~\cite{THSR}. 
Therefore, the next step should go to the study for heavy nuclei within a microscopic cluster model
by using the THSR model, which is most suitable 
for the gas-like cluster structure with a microscopic aspect.
Description of $\alpha$ distribution on the surface or inside the sphere based on this approach is on going,
and we compare with the results of the geometric configurations.

\begin{acknowledgments}
Numerical calculation has been performed using the computer facility of 
Yukawa Institute for Theoretical Physics,
Kyoto University. This work was supported by JSPS KAKENHI Grant Number 17K05440.
\end{acknowledgments}


\begin{references}

\bibitem{PhysRevC97.011301R}
Akihiro Tohsaki and Naoyuki Itagaki,
Phys. Rev. C {\bf 97}, 011301(R) (2018).

\bibitem{Wilson}
H. A. Wilson, Phys. Rev. {\bf 69}, 538 (1948). 

\bibitem{Bulgac}
Aurel Bulgac, Piotr Magierski,
arXiv:nucl-th/0009026 (2000).

\bibitem{Wong}
C. Y. Wong, Ann. Phys. {\bf 77} 279 (1973).

\bibitem{Ichikawa}
T. Ichikawa, J. A. Maruhn, N. Itagaki, K. Matsuyanagi, P.-G. Reinhard, and S. Ohkubo, 
Phys. Rev. Lett. {\bf 109} 232503 (2012). 

\bibitem{Heisenberg}
J. Heisenberg, R. Hofstadter, J. S. McCarthy, I. Sick, B. C. Clark, R. Herman, and D. G. Ravenhall,
Phys. Rev. Lett. {\bf 23}, 1402 (1969).

\bibitem{Brink}
D. M. Brink, in {\it Proceedings of the International School of Physics``Enrico Fermi" Course XXXVI},
edited by C. Bloch (Academic, New York, 1966), 247.


\bibitem{Edmondson}
Amy C. Edmondson,
A Fuller Explanation
-- The Synergetic Geometry of R. Buckminster Fuller, 
(Birkh\"{a}user, Boston, Basel, Stuttgart, 1987).

\bibitem{Kroto}
W.~H.~Kroto, J.~R.~Heath, S.~C.~O'Brien, F.~R.~Curl, R.~E.~C.~Smalley,
Nature, {\bf 318}-6042, 162 (1985).

\bibitem{Kratschmer}
W.~Kr\"{a}tschmer, L.~D.~Lamb, K.~Fostiropoulos and D.~R.~Huffman, Nature 
{\bf 347}, 354 (1990).

\bibitem{Iijima-1}
S. Iijima, Nature {\bf 354}, 56 (1991).

\bibitem{Iijima-2}
S. Iijima and T. Ichihashi, Nature {\bf 363}, 603 (1993).

\bibitem{Tamagaki-1}
R. Tamagaki, Prog. Theor. Phys. Supplement {\bf E68}, 242 (1968).

\bibitem{Tamagaki-2}
R. Tamagaki, Prog. Theor. Phys. {\bf 42}, 748 (1969).

\bibitem{AliBodmer}
S. Ali and A. R. Bodmer, Nucl. Phys. {\bf 80}, 99 (1966).

\bibitem{Satchler}
G. R. Satchler,
L. W. Owen, 
A. J. Elwyn, 
G. L. Morgan, 
and 
R. L. Walter, 
Nucl. Phys. {\bf A112}, 1 (1968).

\bibitem{F1-force}
Akihiro Tohsaki,
Phys. Rev. C {\bf 49}, 1814 (1994).

\bibitem{C12}
N. Itagaki,
Phys. Rev. C {\bf 94}, 064324 (2016). 

\bibitem{O24}
N. Itagaki and A. Tohsaki,
Phys. Rev. C {\bf 97}, 014307 (2018). 

\bibitem{THSR}
A. Tohsaki, H. Horiuchi, P. Schuck and G. R\"{o}pke, Phys.
Rev. Lett. {\bf 87}, 192501 (2001).


\end{references}
\end{document}